\documentstyle[preprint,eqsecnum,aps]{revtex}
\begin{document}
\draft
\title{
\rightline{\rm \normalsize JHU-TIPAC-97001}
\vspace*{-0.4truecm}
\rightline{\rm \normalsize KIAS-P97001}
MAGNETIC PROPERTIES OF NEUTRINOS IN HIGH TEMPERATURE 
$SU(2)_{\scriptscriptstyle{L}}\otimes U(1)$ GAUGE THEORY}
\author{A. Erdas \thanks{erdas@vaxca2.unica.it}}
\address{
Dipartimento di Scienze Fisiche, Universit\`a di Cagliari,
09124 Cagliari, Italy\\
I.N.F.N. Sezione di Cagliari, 09127 Cagliari, Italy}
\author{C. W. Kim \thanks{cwkim@jhuvms.hcf.jhu.edu}}
\address{
Department of Physics and Astronomy,
The Johns Hopkins University, Baltimore, MD 21218 \thanks{permanent address}
\\ School of Physics, 
Korea Institute for Advanced Study, Seoul 130-012, Korea}
\author{T. H. Lee 
\thanks{thlee@plasma.soongsil.ac.kr}}
\address{
Department of Physics,
Soong Sil University, Seoul 156-743, Korea}
\maketitle
%%%%%%%%%%%%%%%%%%%%%%%%%%%%%%%%%%%%%%%%%%%%%%%%%%%%%%%%%%%%%%%%%%%%
\vspace*{-0.4truecm}
\begin {abstract} 
We calculate the finite temperature self-energy for neutrinos 
in the presence of a 
constant magnetic field 
in a medium in the unbroken $SU(2)\otimes
U(1)$ model. We obtain the exact dispersion relation for such neutrinos 
and find that the thermal
effective mass is 
modified by
the magnetic field.
We also find a simple analytic expression for 
the dispersion relation and obtain the index of refraction
for large neutrino momentum.
\end {abstract}
%%%%%%%%%%%%%%%%%%%%%%%%%%%%%%%%%%%%%%%%%%%%%%%%%%%%%%%%%%%%%%%%%%%%%
\section{Introduction}
It has been shown in the past that neutrinos
in the unbroken 
$SU(2)_{\scriptscriptstyle{L}}
\otimes U(1)$ model acquire an effective mass due to finite temperature effects
\cite{Weldon,us,us2}, and it has also been shown that the dispersion
relation for
neutrinos in 
the spontaneously broken Weinberg-Salam model is
significantly changed in the presence of a magnetic field \cite{dolivo}.
It is therefore conceivable that finite temperature 
effects may modify in a significant way the propagation 
of massless Dirac neutrinos in a magnetic field. 
The purpose of 
this paper is to study the propagation of neutrinos in a medium at some 
finite temperature $T$ ($T>250$ GeV) and in the presence of a constant
and homogeneous magnetic field, which in the rest frame of the medium is
$B^\mu=(0,
{\vec B})$ with ${\vec B}$ parallel to the $z$-axis.

The presence of a medium introduces a special Lorentz frame: the center of mass 
of the heat bath. Therefore, in the presence of a magnetic field, 
the neutrino self-energy will be of the form
\begin{equation}
\Sigma(p,B)=\gamma_R(a \not \!p+ b \not \! u +c \not \!\! B
+ b' \not \! u)\gamma_L
\label{1_1}
\end{equation}
where $u_\mu$ is the 4-velocity of the medium, $\gamma_R$ and $\gamma_L$ 
are the right-handed and left-handed projection operators
respectively,
and the coefficients $a$, $b$, $b'$ and 
$c$ are Lorentz-invariant functions.
The two functions $a$ and $b$ have been calculated previously \cite{Weldon}, and
the purpose of our paper is to calculate the functions $b'$ and $c$, which
appear only in the presence of a magnetic field. In particular $b'$
will be explicitly dependent on $B$ and will vanish for $B\rightarrow 0$.

In section II we calculate the finite temperature neutrino self-energy in 
the presence of a constant magnetic field for $T>250$ Gev and $eB\ll T^2$.
In section III we 
obtain the exact neutrino dispersion relation in a medium and in the
presence
of a magnetic field. 
The presence of the magnetic field is shown to 
modify the value of the neutrino thermal effective mass. We also  obtain
a simple analytic expression for the neutrino dispersion relation
in a magnetic field and find the index of refraction
in the case of large neutrino momentum limit.
In section VI we comment on our results and discuss 
implications of our results.
%%%%%%%%%%%%%%%%%%%%%%%%%%%%%%%%%%%%%%%%%%%%%%%%%%%%%%%%%%%%%%%%%%%%%
\section{ Computation of the finite temperature self-energy in a constant
magnetic field}
We start by considering the one loop 
neutrino self-energy $\Sigma_0(p,B)$ in a constant,
homogeneous magnetic field 
${\vec B}$ (which we take to be oriented along the $z$-axis)
at zero temperature. We need to consider 
two Feynman diagrams for this calculation,
the W-lepton diagram and the scalar-lepton diagram. However,  only the 
$W$-lepton diagram is relevant since the scalar-lepton can be neglected 
because the Yukawa coupling $f$ is much smaller than the electroweak
coupling 
$g$. We follow the notation used in Ref. \cite{erdasfeld}, except for taking 
$g^{\mu \nu} = {\rm diag}(+1,-1,-1,-1)$,
and we write the vacuum self-energy as
\begin{equation}
\Sigma_0(x',x'')=i{g^2\over 2}\gamma_R\gamma_\mu
S_-(x',x'')G^\mu_\nu(x',x'')\gamma^\nu \gamma_L
\label{2_0}
\end{equation}
where
\begin{equation}
\gamma_R={{1+\gamma_5}\over 2}\;\;\;,\;\;\;
\gamma_L={{1-\gamma_5}\over 2} \, ,
\end{equation}
$S_-(x',x'')$ is the exact lepton propagator in a constant magnetic field
\cite{schwinger,dittrich} and $G^\mu_\nu(x',x'')$ is the exact 
$W$-propagator in a magnetic field \cite{erdasfeld} and they are given by
\begin{equation}
S_-(x',x'')=i\phi_-(x',x'')\!\int{d^4q\over (2\pi)^4}e^{iq\cdot (x''-x')}
\int_0^\infty \!\!ds_1
{\exp{\left[is_1\left(q^2_{\scriptscriptstyle{\parallel}}+
q^2_{\scriptscriptstyle{\perp}}{\tan z_1\over z_1}\right)\right]}\over
\cos z_1}
\left[
\not\! q_{\scriptscriptstyle{\parallel}}e^{-iz_1\sigma_3}+
{\not\! q_{\scriptscriptstyle{\perp}}\over \cos z_1}
\right]
\label{2_02}
\end{equation}
\begin{equation}
G^\mu_\nu(x',x'')=i\phi_+(x',x'')\int{d^4k\over (2\pi)^4}e^{ik\cdot (x''-x')}
\int_0^\infty ds_2
{\exp{\left[is_2\left(k^2_{\scriptscriptstyle{\parallel}}+
k^2_{\scriptscriptstyle{\perp}}{\tan z_2\over z_2}\right)\right]}\over
\cos z_2}[(\delta^\mu _\nu)_{\scriptscriptstyle{\parallel}}
+(e^{2z_2 \epsilon})^\mu_{{\scriptscriptstyle{\perp}}\nu}]
\label{2_03}
\end{equation}
with
\begin{equation}
\phi_{\pm}(x',x'')=
\exp\left(
\pm i {e\over 2}x''_\mu F^{\mu \nu} x'_\nu
\right)
\label{2_04}
\end{equation}
\begin{equation}
z_i=eBs_i\;, \;\;\;\;i=1,2\,\,;\,\,\,B=|{\vec B}|
\end{equation}
\begin{equation}
\sigma_3=\sigma^{12}={i \over 2}[\gamma^1, \gamma^2]
\end{equation}
where $F^{\mu \nu}$ is the electromagnetic field strength
tensor and for any 4-vector $a^\mu$ we define 
$a^\mu_{\scriptscriptstyle{\parallel}}$ and
$a^\mu_{\scriptscriptstyle{\perp}}$ as
\begin{equation}
a^\mu_{\scriptscriptstyle{\parallel}}=(a^0,0,0,a^3)\;\;\;,\;\;\;
a^\mu_{\scriptscriptstyle{\perp}}=(0,a^1,a^2,0)
\end{equation}
and
\begin{equation}
\left(e^{2z_2 \epsilon}\right)^{\mu}_{{\scriptscriptstyle{\perp}}\nu}=
(\delta^\mu_\nu)_{\scriptscriptstyle{\perp}} \cos 2z_2 
+\epsilon^\mu_\nu \sin 2z_2
\end{equation}
with $\epsilon^1_2=-\epsilon^2_1 =1$ and $\epsilon^\mu_\nu=0$ for all 
other values of the two indices. We are working with the unbroken
version of the Weinberg-Salam theory, and so in 
Eqs.(\ref{2_02}) and 
(\ref{2_03}) we take the $W$-boson mass and the lepton mass to be zero.
We also take the gauge parameter in the $W$-propagator
to be $\xi=1$. Notice that the lepton propagator contains the factor
$\phi_-$, while the $W$-propagator contains $\phi_+$.
This is because 
they are oppositely charged. Substituting the expressions for the propagators
into Eq. (\ref{2_0}), we write the vacuum self-energy in the $\xi=1$ gauge
as \cite{erdasfeld}
\begin{equation}
\Sigma_0(x',x'')=\int{d^4p\over (2\pi)^4}e^{ip\cdot (x''-x')}
\Sigma_0(p,B)
\label{2_08}
\end{equation}
with
\begin{eqnarray}
\Sigma_0(p,B)&=&i{g^2\over 2}\int{d^4q\over (2\pi)^4}
\int{d^4k\over (2\pi)^4} (2\pi)^4 \delta(p-q-k)
\int_0^\infty {ds_1\over
\cos z_1} \int_0^\infty {ds_2\over
\cos z_2}
\nonumber \\
&&{\exp{\left[is_1\left(q^2_{\scriptscriptstyle{\parallel}}+
q^2_{\scriptscriptstyle{\perp}}{\tan z_1\over z_1}\right)\right]}}
{\exp{\left[is_2\left(k^2_{\scriptscriptstyle{\parallel}}+
k^2_{\scriptscriptstyle{\perp}}{\tan z_2\over z_2}\right)\right]}}
\gamma_R\gamma_\mu[(\delta^\mu _\nu)_{\scriptscriptstyle{\parallel}}
+(e^{2z_2 \epsilon})^\mu_{{\scriptscriptstyle{\perp}}\nu}]
\nonumber \\
&&\left[
\not\! q_{\scriptscriptstyle{\parallel}}e^{-iz_1\sigma_3}+
{\not\! q_{\scriptscriptstyle{\perp}}\over \cos z_1}
\right]\gamma^\nu \gamma_L\,\,.
\label{2_1}\end{eqnarray}

When we compute temperature effects, we will be considering temperatures 
for which the $SU(2)_{\scriptscriptstyle{L}}
\otimes U(1)$ symmetry is restored, i.e. $T\ge 250$ GeV. 
Under these assumptions, it is perfectly reasonable to take 
$eB\ll T^2$, 
and so we rewrite Eq. (\ref{2_1}) keeping terms up to order $O(eB)$
\begin{equation}
\Sigma_0(p,B)=-i{g^2\over 2}\int\!{d^4q\over (2\pi)^4}
\int\!{d^4k\over (2\pi)^4} (2\pi)^4 \delta(p-q-k){1\over q^2}
{1\over k^2}\gamma_R\gamma^\mu\!\left[\not\! q-\left({1\over q^2}+
{2\over k^2}\right)eB\sigma_3\not\! q_{\scriptscriptstyle{\parallel}}
\right]\!\gamma_\mu\gamma_L \, ,
\label{2_7}\end{equation}
where we have carried out some of 
straightforward $\gamma$-algebra and two $s$-integrations.
We can write Eq. (\ref{2_7}) as $\Sigma_0(p,B)=\Sigma_0(p)+\Sigma'_0(p,B)$ 
with 
\begin{equation}
\Sigma_0(p)=-i{g^2\over 2}\int{d^4q\over (2\pi)^4}
\int{d^4k\over (2\pi)^4} (2\pi)^4 \delta(p-q-k){1\over q^2}
{1\over k^2}\gamma_R\gamma^\mu\not\! q\gamma_\mu \gamma_L \, ,
\end{equation}
\begin{equation}
\Sigma_0'(p,B)=i{g^2\over 2}\int{d^4q\over (2\pi)^4}
\int{d^4k\over (2\pi)^4} (2\pi)^4 \delta(p-q-k){1\over q^2}
{1\over k^2}\gamma_R\gamma^\mu\left({1\over q^2}+
{2\over k^2}\right)eB\sigma_3\not\! q_{\scriptscriptstyle{\parallel}}
\gamma_\mu\gamma_L \, ,
\end{equation}
where $\Sigma_0(p)$ is the usual neutrino self-energy in vacuum in the absence 
of a magnetic field
and $\Sigma_0'(p,B)$ is the part that depends on $B$. 
The finite temperature
corrections $\Sigma_T(p)$ to $\Sigma_0(p)$ have been calculated \cite{Weldon}
and are well-known, so we focus our attention to the calculation of the temperature corrections to $\Sigma'_0(p,B)$.

We calculate the temperature effects on the neutrino self-energy in a magnetic
field using the imaginary time method of finite temperature field theory
in the rest frame of the 
medium, where the magnetic field is $B^\mu=(0,
{\vec B})$ with ${\vec B}$ pointing in the $z$-direction.
We make the following standard substitutions \cite{kapusta,Dolan} in Eq. 
(\ref{2_7})
\begin{equation}
p^0={\pi i \over \beta}(2n_p+1)\;\;,\;\;\;q^0={\pi i \over \beta}(2n_q+1)\;\;
,\;\;\;k^0={\pi i \over \beta}2n_k
\label{2_8}\end{equation}
\begin{equation}
\int{d^4q\over (2\pi)^4}\rightarrow {i\over \beta} \sum_{n_q}
\int{d^3{\vec q}\over (2\pi)^3}
\;\;,\;\;\;
\int{d^4k\over (2\pi)^4}\rightarrow {i\over \beta} \sum_{n_k}
\int{d^3{\vec k}\over (2\pi)^3}
\end{equation}
\begin{equation}
(2\pi)^4\delta^{(4)}(p-q-k)\rightarrow -i\beta (2\pi)^3
\delta_{n_p,n_q + n_k}\delta^{(3)}({\vec p}-{\vec q}-{\vec k}) \, ,
\end{equation}
where $n_p=n_f(p)$, $n_q=n_f(q)$, and $n_k=n_b(k)$ are the fermion and the boson
occupation numbers(given in Eq.(2.22)),
$\delta_{n_p,n_q + n_k}$ is the Kronecker delta, and $\beta=1/T$. 
After these substitutions, which are the standard procedure for going from
zero temperature to finite temperature \cite{kapusta,Dolan}, we obtain $\Sigma'(p,B)$
which will be separated as
$\Sigma_0'(p,B)+
\Sigma'_T(p,B)$
where $\Sigma_0'$ is the vacuum self-energy and $\Sigma_T'$ is the finite 
temperature correction
\begin{eqnarray}
\Sigma'(p,B)&=&i{g^2\over 2}\int{d^3q\over (2\pi)^3}
\int{d^3k\over (2\pi)^3} (2\pi)^3 \delta({\vec p}-{\vec q}-{\vec k})
\left(i\over \beta \right)^2\sum_{n_q,n_k}(-i\beta)
\delta_{n_p,n_q + n_k}
\nonumber \\
&&\gamma_R\gamma^\mu{\left({1\over q^2}+{2\over k^2}\right)eB
\sigma_3\not\! q_{\scriptscriptstyle{\parallel}}\over q^2 k^2}
\gamma_\mu\gamma_L\, .
\label{2_13}\end{eqnarray}
We first evaluate the double sum
\begin{equation}
I=\left(i\over \beta \right)^2\sum_{n_q,n_k}(-i\beta)
\delta_{n_p,n_q + n_k}
{\left({1\over q^2}+
{2\over k^2}\right)\not\! q_{\scriptscriptstyle{\parallel}}\over 
q^2 k^2}
\label{2_14}\end{equation}
by doing analytic continuations and employing contour integrals. We need to use
an analytic continuation of the Kronecker delta \cite{kapusta}
\begin{equation}
\beta\delta_{n_p,n_q + n_k}={e^{\beta (k^0+q^0)}-
e^{\beta p^0}\over p^0-q^0-k^0} \, ,
\label{2_15}
\end{equation}
since this procedure guarantees that the normal vacuum is recovered in the 
limit of zero temperature \cite{norton}.

By employing Eq. (\ref{2_15}) the double sum $I$ can be evaluated in a closed 
form
\begin{eqnarray}
I&=&-{i\over 4q^2}n'_f(q)\left[{-\gamma^0q+\gamma^3 q^3\over
(p^0-q)^2-k^2}+{\gamma^0q+\gamma^3 q^3\over
(p^0+q)^2-k^2}\right]
\nonumber \\
&&-{i\over 2k^2}\left(n'_b(k)-{n_b(k)\over k}\right)\left[
{\gamma^0(p^0-k)-\gamma^3 q^3\over
(p^0-k)^2-q^2}+{\gamma^0(p^0+k)-\gamma^3 q^3\over
(p^0+k)^2-q^2}\right]
\nonumber \\
&&+{i\over 4q^2}n_f(q)\biggl\{-2(p^0+q){(-\gamma^0q+\gamma^3 q^3)\over
\left[(p^0-q)^2-k^2\right]^2}
+{\gamma^3 q^3\over q}{1\over
(p^0-q)^2-k^2}
\nonumber \\
&&
+2(p^0-q){(\gamma^0q+\gamma^3 q^3)\over
\left[(p^0+q)^2-k^2\right]^2}
+{\gamma^3 q^3\over q}{1\over
(p^0+q)^2-k^2}
\biggr\}
\nonumber \\
&&-{i\over 2k^2}n_b(k)\biggl\{
{\gamma^0\left[(p^0-k)^2+q^2
\right]-2(p^0-k)\gamma^3 q^3 \over\left[(p^0-k)^2-q^2\right]^2}+k
{\gamma^0(p^0+k)- \gamma^3 q^3\over\left[(p^0+k)^2-q^2\right]^2}
\nonumber \\
&&
+k {\gamma^0(p^0-k)- \gamma^3 q^3\over\left[(p^0-k)^2-q^2\right]^2}
-{\gamma^0\left[(p^0+k)^2+q^2\right]-2(p^0+k)\gamma^3 q^3 \over
\left[(p^0+k)^2-q^2\right]^2}
\biggr\}+{\rm vacuum \, \,terms}
\label{2_16}
\end{eqnarray}
where the vacuum terms do not depend on the boson and fermion occupation 
numbers
\begin{equation}
n_b(k)={1\over e^{\beta|{\vec k}|}-1}\;\;\;,\;\;
n_f(q)={1\over e^{\beta|{\vec q}|}+1}
\end{equation}
and we have used the notation $q=|{\vec q}|$, $k=|{\vec k}|$, 
\begin{equation}
n'_b(k)={dn_b(k)\over dk}\,\,\,,\,\,\,n'_f(q)={dn_f(q)\over dq}
\end{equation}
and these notations will be used hereafter in this article.

We now substitute  the expression of $I$ in Eq.(2.21)
into Eq.(\ref{2_13}) for  $\Sigma'(p,B)$ , and then 
perform the ${\vec
q}$-integration. After the 
${\vec q}$-integration is done, we change ${\vec p}-{\vec k}$ to $\vec{k}$
in the
${\vec k}$-integration of the terms proportional to $n_f$ and $n_f'$, 
and obtain
\begin{eqnarray}
\Sigma'_T(p,B)&=&i{g^2\over 2}\int{d^3k\over (2\pi)^3}eB\gamma_R\gamma^\mu
\sigma_3\Biggl\{
-{i\over 2k^2}n'_b(k)\left({{\not \! p_{\scriptscriptstyle{\parallel}}-
\gamma^0k+\gamma^3k^3}\over
A_-}+{{\not \! p_{\scriptscriptstyle{\parallel}}+\gamma^0k+\gamma^3k^3}\over
A_+}\right)
\nonumber \\
&&-{i\over 4k^2}n'_f(k)
\left({{-\gamma^0k+\gamma^3k^3}\over A_-}
+{{\gamma^0k+\gamma^3k^3}\over A_+}
\right)+{i\over 2k^3}n_b(k)\left(
{\not \! p_{\scriptscriptstyle{\parallel}}+\gamma^3k^3}\right)\!\left({1\over
A_-}+{1\over A_+}\right)
\nonumber \\
&&+{i\over 4k^3}n_f(k)\gamma^3k^3\left({1\over
A_-}+{1\over A_+}\right)-{i\over 2k^2}n_b(k)
\Biggl[(2p^0-k){{(\not \! p_{\scriptscriptstyle{\parallel}}-
\gamma^0k+\gamma^3k^3)}\over A_-^2}
\nonumber \\
&&-(2p^0+k){{(\not \! p_{\scriptscriptstyle{\parallel}}+\gamma^0k+
\gamma^3k^3)}\over A_+^2}
\Biggr]-{i\over 2k^2}n_f(k)\Biggl[(p^0+k){{(-\gamma^0k+\gamma^3k^3)}\over
A_-^2}
\nonumber \\
&&-(p^0-k){{(\gamma^0k+\gamma^3k^3)}\over A_+^2}
\Biggr]\Biggr\}\gamma_\mu\gamma_L \, ,
\label{2_19}
\end{eqnarray}
where we have used the notations $p^\mu=(\omega,{\vec p})$,  
$A_{\pm}=\omega^2-|{\vec p}|^2\pm2k\omega+2{\vec k}\cdot{\vec p}$, and
have written $\Sigma'_T$ instead of $\Sigma'$ because the vacuum terms were 
dropped
from the sum $I$. After the integration over the angles and an integration 
by parts, 
Eq.(2.24) reduces to
\begin{eqnarray}
\Sigma'_T(p,B)&=&-{g^2\over 2}\int_0^\infty{k\,d\!k\over(4\pi)^2}{2eB
\gamma_R\gamma^\mu\sigma_3\over \omega^2-|{\vec p}|^2}
\Biggl[\left(\gamma^0+
{\gamma^3p^3\over |{\vec p}|}\right)\omega_-\left(1-{\omega_-
\over |{\vec p}|}\right)
\left({n_b(k)\over \omega^2_+-k^2}-{n_f(k)\over \omega^2_--k^2}\right)
\nonumber \\
&&-\left(-\gamma^0+
{\gamma^3p^3\over |{\vec p}|}\right)\omega_+\left(1+
{\omega_+\over |{\vec p}|}\right)
\left({n_b(k)\over \omega^2_--k^2}-{n_f(k)\over \omega^2_+-k^2}\right)
+{4\over k^2}n_b(k)\not \! p_{\scriptscriptstyle{\parallel}}
\nonumber \\
&&+[n_b(k)-n_f(k)]{\gamma^3p^3\over |{\vec p}|}{\omega^2-|{\vec p}|^2\over 4k
|{\vec p}|^2}
\ln\left({\omega_+-k\over\omega_++k}{\omega_-+k\over\omega_--k}\right)\Biggr]
\gamma_\mu\gamma_L
\label{2_20}
\end{eqnarray}
where, following the notation introduced by Weldon in Ref. \cite{Weldon},
\begin{equation}
\omega_\pm={\omega\pm |{\vec p}|\over 2}\,\,.
\label{2_21}
\end{equation}
Now the last integration is performed by using the following
\begin{equation}
\int_0^\infty{k\,d\!k\over \omega_\pm^2-k^2}n_b(k)=-{1\over 2}\ln\left(
{\beta\omega_\pm\over 2\pi}\right)-{C_E\over 2}+
{\beta^2\omega_\pm^2\over 8\pi^2}\zeta(3)+O(\beta^4 )
\end{equation}
\begin{equation}
\int_0^\infty{k\,d\!k\over \omega_\pm^2-k^2}n_f(k)={1\over 2}\ln\left(
{2\beta\omega_\pm\over \pi}\right)+{C_E\over 2}-
{7\beta^2\omega_\pm^2\over 8\pi^2}\zeta(3)+O(\beta^4 )
\end{equation}
where $C_E=0.5772$ is the Euler-Mascheroni constant,
\begin{equation}
\zeta(3)=\sum_{n=1}^\infty{1\over n^3}.
\end{equation}
In the above, the terms of order $O(\beta^4)$ or higher have been 
neglected because $T>250$ GeV.
We finally obtain the value of the finite temperature self-energy in a 
magnetic field
\begin{eqnarray}
\Sigma'_T(p,B)&=&-{g^2\over (4\pi)^2}{eB\gamma_R\gamma^\mu
\sigma_3\over \omega^2-|{\vec p}|^2}\Biggl\{
2 \not \!p_{\scriptscriptstyle{\parallel}}
\Biggl[-{1\over 2}\ln{\beta^2(\omega^2
-|{\vec p}|^2)\over 4\pi^2}-C_E+
{\beta^2\over 8\pi^2}\zeta (3)(7\omega^2_++7\omega^2_-
\nonumber \\
&&-3\omega_+\omega_-)\Biggr]+\gamma^3p^3{\omega^2-|{\vec p}|^2
\over 4|{\vec p}|^2}\Biggl[{\omega\over |{\vec p}|}
\ln {\omega_-\over \omega_+}
+2-{\beta^2\over 12\pi^2}\zeta (3)(-31\omega^2_+-31\omega^2_-
\nonumber \\
&&+80\omega_+
\omega_-)\Biggr]+4\not \!p_{\scriptscriptstyle{\parallel}}\int_0^\infty
{n_b(k)\over k}dk\Biggl\}\gamma_\mu\gamma_L\,\,.
\label{2_25}
\end{eqnarray}
Notice that the last term in 
Eq.(\ref{2_25}) is an infrared divergence.
It turns out that the infrared divergence will be canceled if we include the 
neutrino bremsstrahlung (soft $Z$ boson emitting process) integrated in a 
range $E<\Delta E$ with a typical energy scale $\Delta E$, as is well-known in
quantum electrodynamics \cite{itzykson}, and therefore we can neglect it.
We can further simplify
the expression of $\Sigma'_T(p,B)$ by using the 
following identity \cite{dolivo}, which is valid for any $4\times 4$ matrix $A$
\begin{equation}
\gamma^\mu\gamma_LA\gamma_\mu\gamma_L=-\left({\rm Tr} A\gamma^\mu\gamma_L
\right)\gamma_\mu\gamma_L
\end{equation}
and find
\begin{eqnarray}
\Sigma'_T(p,B)&=&-{g^2\over (4\pi)^2}\Biggl\{{2 \gamma_R (e\omega\not \!\!B
+e{\vec B}\cdot{\vec p}\not \! u)\gamma_L\over \omega^2-|{\vec p}|^2}
\Biggl[-\ln{\beta^2(\omega^2-|{\vec p}|^2)\over 4\pi^2}-2C_E
\nonumber \\
&&+{\beta^2\over 4\pi^2}\zeta (3)(7\omega^2_++7\omega^2_--3\omega_+
\omega_-)\Biggr]-
{\gamma_R e{\vec B}\cdot{\vec p}
\not \! u\gamma_L\over 2|{\vec p}|^2}
\Biggl[{\omega\over |{\vec p}|}
\ln {\omega_-\over \omega_+}
\nonumber \\
&&+2-{\beta^2\over 12\pi^2}\zeta (3)(-31\omega^2_+-31\omega^2_-
+80\omega_+
\omega_-)\Biggr]
\Biggr\}
\label{2_27}
\end{eqnarray}
where we have used $Bp^3={\vec B}\cdot{\vec p}$, 
$B\gamma_3= \,\,\not \!\!\!B$ and $\gamma_0=\,\not\! u$, 
which are true in the rest frame of the medium.
It is clear that for very high temperature (i.e. $\beta \ll 1$)
the term proportional to $\ln \beta^2$ is dominant compared with 
terms proportional to $\beta^2$ or independent of $\beta$, and 
therfore the first logarithmic term 
in Eq. (\ref{2_27}) is 
dominant, and we obtain
\begin{equation}
\Sigma'_T(p,B)
\simeq {2g^2\over (4\pi)^2}{ \gamma_R (e\omega\not \!\!B
+e{\vec B}\cdot{\vec p}\not \! u)\gamma_L\over \omega^2-|{\vec p}|^2}
\ln{\beta^2(\omega^2-|{\vec p}|^2)\over 4\pi^2}
\label{2_28}
\end{equation}
%%%%%%%%%%%%%%%%%%%%%%%%%%%%%%%%%%%%%%%%%%%%%%%%%%%%%%%%%%%%%%%%%%%%%
\section{ Dispersion relation in matter in the presence of a magnetic field}

In the previous section we have calculated the $B$-dependent part of the finite
temperature neutrino self-energy in a medium in the presence of an external
magnetic field. We now cast the result in a form given in Eq.(1.1)
\begin{equation}
\Sigma_T(p,B)=\gamma_R(a \not \!p+ b \not \! u +c \not \!\! B
+ b' \not \! u)\gamma_L
\label{3_1}
\end{equation}
where $u_\mu$ is the 4-velocity of the medium (in the rest frame of the
medium $u_\mu=(1,0)$), and the coefficients $a$ and $b$ are well known
\cite{Weldon}
\begin{equation}
a={M^2\over |{\vec p}|^2}\left(1-{\omega\over 2|{\vec p}|}\ln {\omega_+\over
 \omega_-}\right)
\label{3_2}
\end{equation}
\begin{equation}
b={M^2\over |{\vec p}|}\left[-{\omega\over |{\vec p}|}+
\left({\omega^2\over |{\vec p}|^2}-1\right){1\over 2}\ln {\omega_+\over
 \omega_-}\right]
\label{3_3}
\end{equation}
with $\omega_\pm$ defined 
in Eq.(\ref{2_21}) and the thermal
effective mass $M$ for neutrinos 
in the unbroken 
$SU(2)_{\scriptscriptstyle{L}}\otimes U(1)$ model given by 
$M^2={3\over 32}g^2T^2$.
The other two coefficients $b'$ and $c$ are immediately obtained 
from Eq.(\ref{2_28}) of our calculation as
\begin{equation}
b'=-{g^2\over 8\pi^2}{e{\vec B}\cdot{\vec p} \over \omega^2-|{\vec p}|^2}
\ln{(2\pi T)^2\over \omega^2-|{\vec p}|^2}
\label{3_4}
\end{equation}
\begin{equation}
c=-{g^2\over 8\pi^2}{e\omega \over \omega^2-|{\vec p}|^2}
\ln{(2\pi T)^2\over \omega^2-|{\vec p}|^2}\,\,.
\label{3_5}
\end{equation}
Starting from the inverse of the full neutrino propagator in matter in
a magnetic field $S^{-1}=\gamma_R\!\not\!\!p \,\,\,\gamma_L + \Sigma_T(p,B)$,
with $\Sigma_T(p,B)$ given by 
Eq.(\ref{3_1}), and 
following the same procedure as in Refs. \cite{Weldon,us,us2}, we square
$S^{-1}$ and set it equal to zero to find
\begin{equation}
[\omega(1+a)+b+b']^2=|{\vec p}(1+a)+c{\vec B}|^2
\label{3_6}
\end{equation}
from which the dispersion relation is easily obtained.
The positive-energy solution of Eq. (\ref{3_6}) occurs when $\omega$ and 
${\vec p}$ satisfy
\begin{equation}
\omega(1+a)+b=\left[(1+a)^2|{\vec p}|^2+c^2B^2-
2bb'-b'^2\right]^{1/2}\,\,.
\label{3_7}
\end{equation}
The above equation is easily derived from 
Eq.(\ref{3_6}) using the relation
$\omega b'=c{\vec p}\cdot{\vec B}$ and is the exact finite temperature
dispersion relation
for neutrinos in a magnetic field. The last three terms on the right-hand
side of 
Eq.(\ref{3_7}) are due to the presence of the magnetic
field, and vanish for $B\rightarrow 0$. 

Now we analyze the effect of the 
magnetic field on the neutrino thermal effective mass and obtain a simple 
analytic form of the dispersion relation for 
$|{\vec p}|\gg M$. Weldon \cite{Weldon} showed that 
the solution of the thermal dispersion relation with $|{\vec p}|=0$
is not $\omega=0$, but
$\omega=M$, proving that $M$ is an effective mass due to thermal
effects.
We take 
Eq.(\ref{3_7})
in the limit for ${\vec p}\rightarrow 0$, and obtain
\begin{equation}
\omega -{M^2\over \omega}={g^2\over 4\pi^2}{eB\over\omega}
\ln{2\pi T\over\omega}
\label{3_8}
\end{equation}
whence after solving for $\omega$ with
the assumption that $eB\ll T^2$,
we find 
\begin{equation}
\omega =M+{g^2\over 8\pi^2}{eB\over M}
\ln{2\pi T\over M}\,\,.
\label{3_9}
\end{equation}
This shows that the thermal effective mass $M$ of the neutrino
is modified, in the presence of a magnetic field, to
\begin{equation}
M_B= M+{g^2\over 8\pi^2}{eB\over M}
\ln{2\pi T\over M}.
\label{3_10}
\end{equation}
The analytic form of the thermal dispersion relation without magnetic field
was obtained by Weldon
\cite{Weldon}
\begin{equation}
\omega =|{\vec p}|+{M^2\over |{\vec p}|}-{M^4\over 2
|{\vec p}|^3}
\ln\left({2|{\vec p}|^2\over M^2}\right)
+\cdots\;\;\;\;\;
(|{\vec p}|\gg M) \, .
\label{3_12}
\end{equation}
In this article with a magnetic field 
we consider only the case $|{\vec p}|\gg M$ because
$T\ge 250$ GeV.
Solving Eq.(\ref{3_7}) for 
$|{\vec p}|\gg M$, we find the corrections to 
Eq.(\ref{3_12}) due to the presence of a magnetic field
\begin{equation}
\omega =|{\vec p}|+
{M^2\over |{\vec p}|}-
{M^4\over 2
|{\vec p}|^3}\ln\left({2|{\vec p}|^2\over M^2}\right)
-{g^2\over 4\pi ^2}{e{\vec B}\cdot{\hat p}\over |{\vec p}|}
\ln \left({{\sqrt{2}}\pi T\over M}\right)+
\cdots\;\;\;\;\;
(|{\vec p}|\gg M)
\label{3_14}
\end{equation}
where ${\hat p}={\vec p}/|{\vec p}|$ and the term proportional
to ${\vec B}\cdot{\hat p}$ is the leading order magnetic field
correction. From 
Eq.(\ref{3_14}) we obtain the
index of refraction $n=|{\vec p}|/\omega$ for neutrino in a
magnetic field
\begin{equation}
n= 1-
{M^2\over |{\vec p}|^2}+{g^2\over 4\pi ^2}
{e{\vec B}\cdot{\hat p}\over |{\vec p}|^2}
\ln \left({{\sqrt{2}}\pi T\over M}\right)+
\cdots\;\;\;\;\;
(|{\vec p}|\gg M)\,\,.
\label{3_16}
\end{equation}
It is very interesting to notice that for neutrinos 
moving in different directions we obtain different dispersion 
relations and therefore different speeds. 
%%%%%%%%%%%%%%%%%%%%%%%%%%%%%%%%%%%%%%%%%%%%%%%%%%%%%%%%%%%%%%%%%%%%%
\section{ Discussion}

We have shown that in the presence of a magnetic field $eB\ll T^2$, the
thermal 
effective mass $M$ for neutrinos in the unbroken Weinberg-Salam model
increases by
\begin{equation}
{g^2\over 8\pi^2}{eB\over M}
\ln{2\pi T\over M}.
\label{4_1}
\end{equation}
The thermal dispersion relation is also modified by 
\begin{equation}
-{g^2\over 4\pi ^2}{e{\vec B}\cdot{\hat p}\over |{\vec p}|}
\ln \left({{\sqrt{2}}\pi T\over M}\right) \, ,
\;\;\;\;\;
(|{\vec p}|\gg M)
\label{4_3}
\end{equation}
which is the leading order magnetic field corrections and the index of
refraction is
\begin{equation}
n= 1-
{M^2\over |{\vec p}|^2}+{g^2\over 4\pi ^2}
{e{\vec B}\cdot{\hat p}\over |{\vec p}|^2}
\ln \left({{\sqrt{2}}\pi T\over M}\right)+
\cdots\;\;\;\;\;
(|{\vec p}|\gg M)\,\,.
\label{4_5}
\end{equation}
The above results 
show that neutrinos propagating in different directions have different 
dispersion relations. Our findings might have 
interesting applications
in the field of the early cosmology, since our calculations are valid for
$eB\ll T^2$, which implies magnetic fields $B\ll 10^{24}$Gauss, and
therefore primordial magnetic fields in the very early universe
might have affected the propagation of neutrinos in the way we have
calculated.
%%%%%%%%%%%%%%%%%%%%%%%%%%%%%%%%%%%%%%%%%%%%%%%%%%%%%%%%%%%%%%%%%%%%%
\acknowledgements
We wish to thank Gordon Feldman for helpful discussions.
A. Erdas wishes to thank the High Energy Theory Group of 
the Johns Hopkins University for the hospitality extended to him
during his several visits. This work is supported in part by the 
Basic Science Research Institute Program, Ministry of Education, 
Korea, 1996, 
Project No. BSRI-96-2418, and in part by Soong Sil University (THL).
%%%%%%%%%%%%%%%%%%%%%%%%%%%%%%%%%%%%%%%%%%%%%%%%%%%%%%%%%%%%%%%%%%%%%

%%%%%%%%%%%%%%%%%%%%%%%%%%%%%%%%%%%%%%%%%%%%%%%%%%%%%%%%%%%%%%%%%%%%%
\end{document}